\documentclass{PoS}

\title{Relic Right-handed Dirac Neutrinos and Cosmic Neutrino Background}

\ShortTitle{Right-handed Neutrinos and Cosmic Neutrino Background}

\author{\speaker{Shun Zhou}\thanks{This work was supported in part by the National Recruitment Program for Young Professionals and by the CAS Center for Excellence in Particle Physics (CCEPP).}\\
        Institute of High Energy Physics, Chinese Academy of Sciences, Beijing 100049, China\\
        Center for High Energy Physics, Peking University, Beijing 100871, China\\
        E-mail: \email{zhoush@ihep.ac.cn}}

\abstract{The PTOLEMY experiment, implementing a $100~{\rm g}$ surface-deposition tritium target, is promising to detect cosmic neutrino background via $\nu^{}_e + {^3}{\rm H} \to {^3}{\rm He} + e^-$. In this talk, I consider a thermal production of right-handed Dirac neutrinos in the early Universe, and investigate their impact on the capture rate of cosmic relic neutrinos at PTOLEMY.}

\FullConference{Neutrino Oscillation Workshop\\
		4 - 11 September, 2016\\
		Otranto (Lecce, Italy)}

\begin{document}

\section{Introduction}

One milestone achievement of the big bang cosmology is the prediction for cosmic microwave background (CMB), which has now been precisely measured and led to a tremendous progress in our understanding of the Universe~\cite{Weinberg:2008zzc}. As another solid prediction from the big bang theory, cosmic neutrino background (C$\nu$B) should exist as well and it must carry useful information about the early Universe when it was just one second old. Therefore, a direct detection of C$\nu$B in terrestrial laboratories is of crucial importance to test the standard cosmology on the one hand, and to open a new window on probing intrinsic properties of neutrinos themselves on the other hand.

When the temperature of the Universe dropped down to $T = T^{}_{\rm L} \approx 1~{\rm MeV}$, the Hubble expansion rate exceeded the weak interaction rate of left-handed neutrinos $\nu^{}_{\rm L}$ and right-handed antineutrinos $\overline{\nu}^{}_{\rm R}$, and thus both $\nu^{}_{\rm L}$ and $\overline{\nu}^{}_{\rm R}$ decoupled from the thermal bath. At this moment, $\nu^{}_{\rm L}$ and $\overline{\nu}^{}_{\rm R}$ were extremely relativistic, given neutrino masses $m^{}_\nu \lesssim 0.1~{\rm eV}$~\cite{pdg}. Consequently, the number density $n^{}_{\nu^{}_{\rm l}}$ of left-helical neutrinos $\nu^{}_{\rm l}$ was equal to that $n^{}_{\nu^{}_{\rm L}}$ of left-handed neutrinos $\nu^{}_{\rm L}$, while the number density $n^{}_{\nu^{}_{\rm r}}$ of right-helical neutrinos $\nu^{}_{\rm r}$ is vanishing. Hence we have $n^{}_{\nu^{}_{\rm l}} = n^{}_{\nu^{}_{\rm L}}$ and $n^{}_{\nu^{}_{\rm r}} = 0$ for neutrinos, while $n^{}_{\overline{\nu}^{}_{\rm r}} = n^{}_{\overline{\nu}^{}_{\rm R}}$ and $n^{}_{\overline{\nu}^{}_{\rm l}} = 0$ for antineutrinos, at the decoupling temperature $T^{}_{\rm L}$. Since the helicity operator commutes with the free Hamiltonian, neutrino helicities after decoupling are always conserved in the rest frame of C$\nu$B. As the Universe is expanding, the neutrino temperature will be red-shifted. Nowadays, the average temperature of CMB photons is $T^{}_\gamma = 2.725~{\rm K}$, which is
related to the neutrino temperature $T^{}_\nu = (4/11)^{1/3} T^{}_\gamma \approx 1.945~{\rm K}$. The difference between $T^{}_\gamma$ and $T^{}_\nu$ can be traced back to the reheating of photons via $e^+ e^- \to \gamma \gamma$ around $T = 0.5~{\rm MeV}$. Therefore, we obtain average number densities $\overline{n}^{}_{\nu^{}_{\rm l}} = \overline{n}^{}_{\overline{\nu}^{}_{\rm r}} \approx 56~{\rm cm}^{-3}$ per neutrino flavor in the present Universe.

It is a great challenge to detect such low-energy relic neutrinos, whose average momentum is $\langle p^{}_\nu \rangle \approx 5.28\times 10^{-4}~{\rm eV}$. One promising approach is to seize non-relativistic relic neutrinos by radioactive $\beta$-decaying nuclei~\cite{Weinberg:1962zza}, e.g., $\nu^{}_e + {^3{\rm H}} \to {^3{\rm He}} + e^-$, for which there is no energy threshold of $\nu^{}_e$. In this process, the signal is simply a peak located at a distance of $2m^{}_\nu$ from the endpoint of the $\beta$ spectrum for ${^3{\rm H}} \to {^3{\rm He}} + \overline{\nu}^{}_e + e^-$~\cite{Irvine:1983nr}. The recently proposed PTOLEMY experiment will implement a 100 g surface-deposition tritium target and could reach an energy resolution of $0.15~{\rm eV}$, which will hopefully discover C$\nu$B~\cite{Betts:2013uya}. See, e.g., Refs.~\cite{Ringwald:2005zf,Vogel:2015vfa}, for a review on this topic.

\section{Dirac Neutrinos}

The simplest extension of the Standard Model (SM) to accommodate tiny neutrino masses is to add three right-handed neutrino singlets and generate Dirac masses for neutrinos in the same way as for quarks and charged leptons. However, the huge hierarchy between neutrino masses $m^{}_\nu \lesssim 0.1~{\rm eV}$ and top-quark mass $m^{}_t = 1.71 \times 10^{12}~{\rm eV}$ needs to be further explained. Since the Yukawa couplings of Dirac neutrinos are extremely small $y^{}_\nu \lesssim 10^{-12}$, the direct production of right-handed neutrinos $\nu^{}_{\rm R}$ and left-handed antineutrinos $\overline{\nu}^{}_{\rm L}$ in the early Universe is highly suppressed~\cite{Antonelli:1981eg,Zhang:2015wua}. Therefore, we have $n^{}_{\nu^{}_{\rm r}} = n^{}_{\overline{\nu}^{}_{\rm l}} = 0$ at $T = T^{}_{\rm L}$ and today as well.

\begin{figure}[t!]
\centering
\includegraphics[width=4.0in]{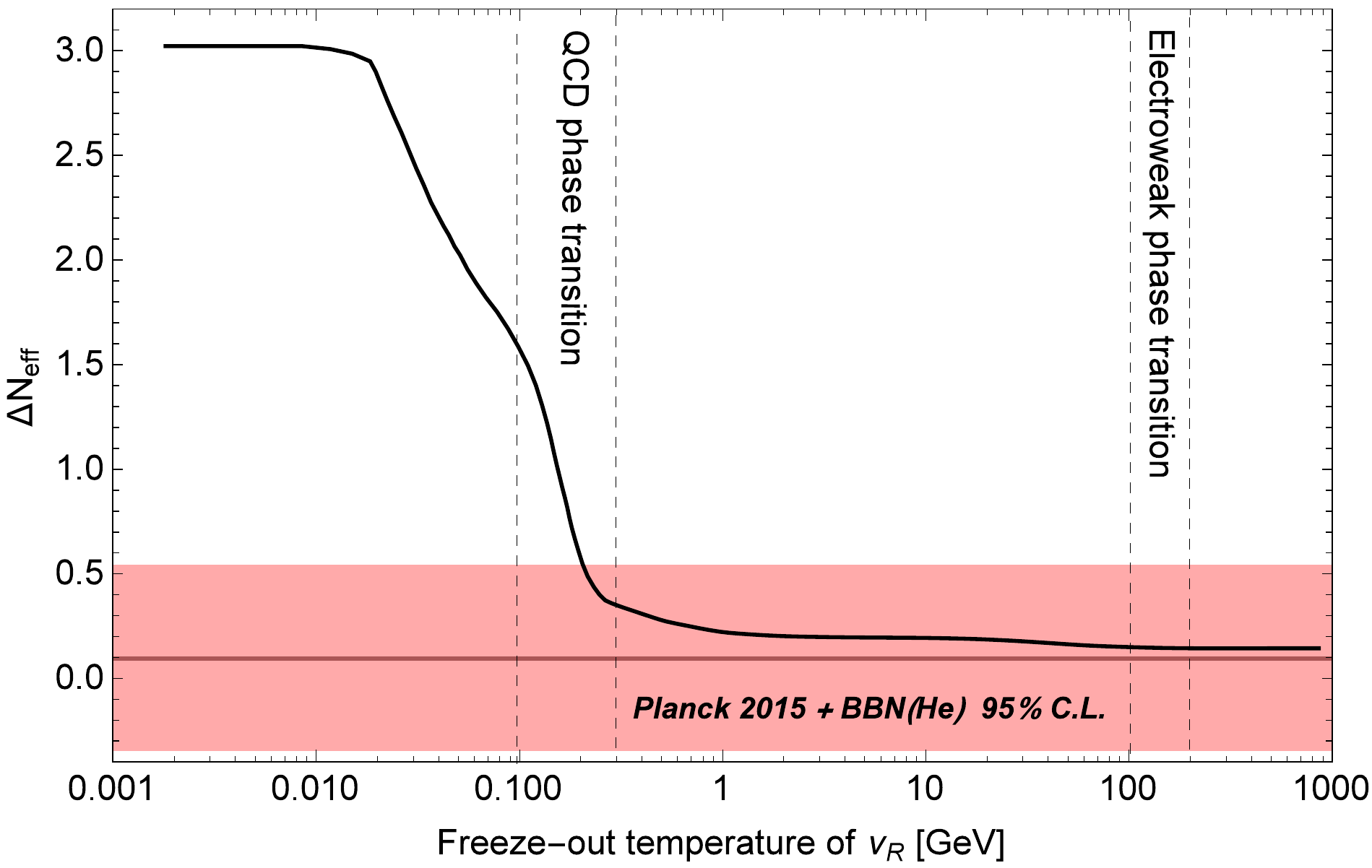}
\caption{The extra effective number of neutrinos $\Delta N^{}_{\rm eff}$ is shown with respect to the decoupling temperature $T^{}_{\rm R}$ of right-handed neutrinos~\cite{Zhang:2015wua}.}
\label{fig:DeltaN}
\end{figure}

In Ref.~\cite{Zhang:2015wua}, a working example has been given to thermally produce right-handed neutrinos $\nu^{}_{\rm R}$ and left-handed antineutrinos $\overline{\nu}^{}_{\rm L}$. In this scenario, primordial magnetic fields $B^{}_0 \approx 10^{24}~{\rm G}$ within a domain size $L^{}_0 > 10^{-7}~{\rm cm}$ are assumed to be generated during the electroweak phase transition at $T = 100~{\rm GeV}$. Although the evolution of such magnetic fields in the early Universe is not yet quite clear, some phenomenological models are available~\cite{Enqvist:1994mb}. It can be shown that massive Dirac neutrinos with a small magnetic dipole moment $\mu^{}_\nu = 3\times 10^{-20}~(m^{}_\nu/0.1~{\rm eV})~\mu^{}_{\rm B}$, where $\mu^{}_{\rm B}$ is the Bohr magneton, can experience spin-flipping conversions $\nu^{}_{\rm L} \to \nu^{}_{\rm R}$ and $\overline{\nu}^{}_{\rm R} \to \overline{\nu}^{}_{\rm L}$ in magnetic fields. For $B^{}_0 \approx 10^{24}~{\rm G}$ and $L^{}_0 > 10^{-7}~{\rm cm}$, these conversions are sufficiently rapid but become out of equilibrium in the epoch of QCD phase transition around $T \approx 200~{\rm MeV}$. As these additional thermal relics contribute to the total energy density just like ordinary neutrinos, they are subject to the cosmological upper bound on the extra effective number of neutrinos, namely, $\Delta N^{}_{\rm eff} < 0.53$ at the $95\%$ confidence level. In Fig.~\ref{fig:DeltaN}, one can observe that the decoupling temperature $T^{}_{\rm R}$ of $\nu^{}_{\rm R}$ and $\overline{\nu}^{}_{\rm L}$ above $200~{\rm MeV}$ is compatible with the cosmological bound.

We should calculate the number densities of $\nu^{}_{\rm r}$ and $\overline{\nu}^{}_{\rm l}$ at present by assuming that the upper bound $\Delta N^{}_{\rm eff} < 0.53$ is saturated. First, it is straightforward to find the number density at $T^{}_{\rm L}$~\cite{Zhang:2015wua}
\begin{eqnarray}
\frac{n^{}_{\nu^{}_{\rm r}}(T^{}_{\rm L})}{n^{}_{\nu^{}_{\rm l}}(T^{}_{\rm L})} = \frac{n^{}_{\nu^{}_{\rm r}}(T^{}_{\rm L})}{n^{}_{\nu^{}_{\rm r}}(T^{}_{\rm R})} \cdot \frac{n^{}_{\nu^{}_{\rm l}}(T^{}_{\rm R})}{n^{}_{\nu^{}_{\rm l}}(T^{}_{\rm L})} = \frac{g^{}_{*{\rm s}}(T^{}_{\rm L})}{g^{}_{*{\rm s}}(T^{}_{\rm R})} \; ,
\end{eqnarray}
where $n^{}_{\nu^{}_{\rm r}}(T^{}_{\rm R}) = n^{}_{\nu^{}_{\rm l}}(T^{}_{\rm R})$ and $n^{}_{\nu^{}_{\rm l}}(T^{}_{\rm R})/n^{}_{\nu^{}_{\rm l}}(T^{}_{\rm L}) = T^3_{\rm R}/T^3_{\rm L}$ hold for neutrinos in thermal equilibrium. For the decoupled $\nu^{}_{\rm R}$ in the adiabatically expanding Universe, the entropy conservation gives rise to $n^{}_{\nu^{}_{\rm r}}(T^{}_{\rm R})/n^{}_{\nu^{}_{\rm r}}(T^{}_{\rm L}) = [g^{}_{*{\rm s}}(T^{}_{\rm R}) T^3_{\rm R}]/[g^{}_{*{\rm s}}(T^{}_{\rm L}) T^3_{\rm L}]$, where $g^{}_{*{\rm s}}$ denotes the effective number of degrees of freedom contributing to the entropy density. Given $T^{}_{\rm R} \approx 200~{\rm MeV}$ and $T^{}_{\rm L} \approx 1~{\rm MeV}$, we get $g^{}_{*{\rm s}}(T^{}_{\rm R}) \approx 38.4$ and $g^{}_{*{\rm s}}(T^{}_{\rm L}) \approx 10.75$, implying that $n^{}_{\nu^{}_{\rm r}}/n^{}_{\nu^{}_{\rm l}} \approx 28\%$, which remains to be constant until today as both $\nu^{}_{\rm r}$ and $\nu^{}_{\rm l}$ are decoupled below $T^{}_{\rm L}$. Thus, the average number densities are $\overline{n}^{}_{\nu^{}_{\rm r}} = \overline{n}^{}_{\overline{\nu}^{}_{\rm l}} \approx 16~{\rm cm}^{-3}$ per neutrino flavor, which should be compared with $\overline{n}^{}_{\nu^{}_{\rm r}} = \overline{n}^{}_{\overline{\nu}^{}_{\rm l}} \approx 0$ in the case without thermal production of $\nu^{}_{\rm R}$ and $\overline{\nu}^{}_{\rm L}$.

\section{Capture Rates}

Now that the C$\nu$B is made of all four helical neutrino states, namely, $\nu^{}_{\rm l}$ and $\overline{\nu}^{}_{\rm r}$ of an average number density $56~{\rm cm}^{-3}$, and $\nu^{}_{\rm r}$ and $\overline{\nu}^{}_{\rm l}$ of $16~{\rm cm}^{-3}$, their capture rate on the tritium target should be changed. The capture rate for $\nu^{}_e + {^3{\rm H}} \to {^3{\rm He}} + e^-$ was first calculated in Ref.~\cite{Cocco:2007za}, and later corrected in Ref.~\cite{Long:2014zva}. Considering an unpolarized tritium target and a neutrino mass eigenstate $\nu^{}_i$ of spin $s^{}_\nu$ (i.e., $+1/2$ or $-1/2$), one can find that the product of the cross section $\sigma^{}_i(s^{}_\nu)$ and the neutrino velocity $v^{}_{\nu^{}_i}$ can be written as $\sigma^{}_i(s^{}_\nu) v^{}_{\nu^{}_i} = {\cal A}(s^{}_\nu) |U^{}_{ei}|^2 \overline{\sigma}$, where $\overline{\sigma} \approx 3.8\times 10^{-45}~{\rm cm}^2$, ${\cal A}(s^{}_\nu) \equiv 1-2s^{}_\nu v^{}_{\nu^{}_i}$ and $U$ is the unitary lepton flavor mixing matrix. For non-relativistic C$\nu$B neutrinos with $v^{}_{\nu^{}_i} \to 0$, we have ${\cal A}(+1/2) = {\cal A}(-1/2) \approx 1$, implying that both left- and right-helical neutrino states can equally be captured~\cite{Long:2014zva}. The total capture rate is then given by
\begin{eqnarray}
\Gamma^{}_{\rm D} = N^{}_{\rm T} \sum^3_{i = 1} \left[\sigma^{}_i(-1/2) v^{}_{\nu^{}_i} \overline{n}^{}_{\nu^{}_{\rm l}} + \sigma^{}_i(+1/2) v^{}_{\nu^{}_i} \overline{n}^{}_{\nu^{}_{\rm r}}\right] \approx N^{}_{\rm T} \overline{\sigma} \left(\overline{n}^{}_{\nu^{}_{\rm l}} + \overline{n}^{}_{\nu^{}_{\rm r}}\right) \; ,
\end{eqnarray}
where $N^{}_{\rm T}$ is the number of tritium nuclei and the unitarity condition $\sum_i |U^{}_{e i}| = 1$ has been used. It is easy to observe that $\Gamma^{}_{\rm D} \approx 4~{\rm yr}^{-1}$ in the standard case~\cite{Long:2014zva} will be increased to $\Gamma^{}_{\rm D} \approx 5.1~{\rm yr}^{-1}$ in the presence of right-handed neutrinos in the early Universe~\cite{Zhang:2015wua}. As pointed out in Ref.~\cite{Long:2014zva}, if massive neutrinos are Majorana particles, both $\nu^{}_{\rm l}$ and $\overline{\nu}^{}_{\rm r}$ (now should be identified as $\nu^{}_{\rm r}$) will participate in the capture process, leading to a twice larger rate $\Gamma^{}_{\rm M} \approx 8~{\rm yr}^{-1}$.

A final remark is about further considerations on C$\nu$B. In Ref.~\cite{Chen:2015dka}, a nonthermal production of $\nu^{}_{\rm R}$ and $\overline{\nu}^{}_{\rm L}$ from inflaton decays has been proposed for Dirac neutrinos. In this scenario, saturating the bound $\Delta N^{}_{\rm eff} < 0.53$, the average number density $\overline{n}^{}_{\nu^{}_{\rm r}} \approx 29~{\rm cm}^{-3}$ and thus a capture rate of $\Gamma^{}_{\rm D} \approx 6.1~{\rm yr}^{-1}$ can be reached. Possible discrimination between thermal and nonthermal spectra of right-handed neutrinos may be achieved by observing the annual modulation at PTOLEMY~\cite{Huang:2016qmh}.

\section*{Acknowledgements}

I would like to thank Guo-yuan Huang and Jue Zhang for enjoyable collaborations on cosmic relic neutrinos, and the organisers of NOW 2016 for kind invitation and warm hospitality.

\end{document}